\documentclass[11pt, letterpaper, onecolumn]{IEEEtran}

\usepackage[ colorlinks = true,
             linkcolor = blue,
             urlcolor  = blue,
             citecolor = red,
             anchorcolor = green,
]{hyperref}

\usepackage[mathscr]{eucal}
\usepackage[cmex10]{amsmath}
\usepackage{epsfig,epsf,psfrag,overpic}
\usepackage{amssymb,amsmath,amsthm,amsfonts,latexsym}
\usepackage{amsmath,graphicx,bm,xcolor,url}
\usepackage[caption=false]{subfig} 
\usepackage{fixltx2e}
\usepackage{array}
\usepackage{verbatim}
\usepackage{bm}
\usepackage{algorithmic, cite}
\usepackage{algorithm}
\usepackage{verbatim}
\usepackage{textcomp}
\usepackage{mathrsfs}
\usepackage{epstopdf}

\catcode`~=11 \def\UrlSpecials{\do\~{\kern -.15em\lower .7ex\hbox{~}\kern .04em}} \catcode`~=13 

\allowdisplaybreaks[1]
 
\newcommand{\nn}{\nonumber}

\newcommand{\calD}{\mathcal{D}}
\newcommand{\calE}{\mathcal{E}}
\newcommand{\calF}{\mathcal{F}}
\newcommand{\calG}{\mathcal{G}}

\newcommand{\calN}{\mathcal{N}}

\newcommand{\calP}{\mathcal{P}}

\newcommand{\bI}{\mathbf{I}}

\newcommand{\bx}{\mathbf{x}}
\newcommand{\bX}{\mathbf{X}}
\newcommand{\by}{\mathbf{y}}
\newcommand{\bY}{\mathbf{Y}}
\newcommand{\bz}{\mathbf{z}}
\newcommand{\bZ}{\mathbf{Z}}

\newcommand{\rmd}{\mathrm{d}}

\newcommand{\rme}{\mathrm{e}}

\newcommand{\rmP}{\mathrm{P}}

\newcommand{\bbE}{\mathbb{E}}

\newcommand{\bbN}{\mathbb{N}}

\newcommand{\bbR}{\mathbb{R}}

\DeclareMathAlphabet{\mathbsf}{OT1}{cmss}{bx}{n}
\DeclareMathAlphabet{\mathssf}{OT1}{cmss}{m}{sl}

\newcommand{\rvC}{\mathsf{C}}

\newcommand{\rvT}{\mathsf{T}}

\newcommand{\rvV}{\mathsf{V}}

\DeclareSymbolFont{bsfletters}{OT1}{cmss}{bx}{n}  
\DeclareSymbolFont{ssfletters}{OT1}{cmss}{m}{n}
\DeclareMathSymbol{\bsfGamma}{0}{bsfletters}{'000}
\DeclareMathSymbol{\ssfGamma}{0}{ssfletters}{'000}
\DeclareMathSymbol{\bsfDelta}{0}{bsfletters}{'001}
\DeclareMathSymbol{\ssfDelta}{0}{ssfletters}{'001}
\DeclareMathSymbol{\bsfTheta}{0}{bsfletters}{'002}
\DeclareMathSymbol{\ssfTheta}{0}{ssfletters}{'002}
\DeclareMathSymbol{\bsfLambda}{0}{bsfletters}{'003}
\DeclareMathSymbol{\ssfLambda}{0}{ssfletters}{'003}
\DeclareMathSymbol{\bsfXi}{0}{bsfletters}{'004}
\DeclareMathSymbol{\ssfXi}{0}{ssfletters}{'004}
\DeclareMathSymbol{\bsfPi}{0}{bsfletters}{'005}
\DeclareMathSymbol{\ssfPi}{0}{ssfletters}{'005}
\DeclareMathSymbol{\bsfSigma}{0}{bsfletters}{'006}
\DeclareMathSymbol{\ssfSigma}{0}{ssfletters}{'006}
\DeclareMathSymbol{\bsfUpsilon}{0}{bsfletters}{'007}
\DeclareMathSymbol{\ssfUpsilon}{0}{ssfletters}{'007}
\DeclareMathSymbol{\bsfPhi}{0}{bsfletters}{'010}
\DeclareMathSymbol{\ssfPhi}{0}{ssfletters}{'010}
\DeclareMathSymbol{\bsfPsi}{0}{bsfletters}{'011}
\DeclareMathSymbol{\ssfPsi}{0}{ssfletters}{'011}
\DeclareMathSymbol{\bsfOmega}{0}{bsfletters}{'012}
\DeclareMathSymbol{\ssfOmega}{0}{ssfletters}{'012}

\newcommand{\tilm}{\tilde{m}}

\newcommand{\tilR}{\tilde{R}}

\newcommand{\eps}{\varepsilon}

\DeclareMathOperator{\var}{\mathsf{Var}}

\newcommand{\bzero}{\mathbf{0}}
\newcommand{\bone}{\mathbf{1}}

\newtheorem{theorem}{Theorem} 
\newtheorem{lemma}[theorem]{Lemma}

\newcommand{\qednew}{\nobreak \ifvmode \relax \else
      \ifdim\lastskip<1.5em \hskip-\lastskip
      \hskip1.5em plus0em minus0.5em \fi \nobreak
      \vrule height0.75em width0.5em depth0.25em\fi}

\begin{document}
\flushbottom

\title{The Third-Order Term in the Normal Approximation for the AWGN  Channel }
 
 \author{Vincent Y.~F.\ Tan$^\dagger$  \hspace{.025in}  and \hspace{.025in}  Marco Tomamichel$^*$   \thanks{$^\dagger$  Department of Electrical and Computer Engineering (ECE), National University of Singapore (NUS) and Institute for Infocomm Research (I$^2$R), Agency for Science, Technology and Research (A*STAR)     (Email:
  \url{vtan@nus.edu.sg})}    \thanks{$^*$ Center for Quantum Technologies, National University of Singapore (Email: \url{cqtmarco@nus.edu.sg})}   }
   

\maketitle

\begin{abstract}
This paper shows that, under the average error probability formalism,  the third-order term in the normal approximation for the additive white Gaussian noise channel with a maximal or equal power constraint  is at least  $\frac{1}{2}\log n + O(1)$. This matches the upper bound derived by   Polyanskiy-Poor-Verd\'u (2010). 
\end{abstract}


\section{Introduction}
The most important continuous alphabet channel in communication systems is the discrete-time additive white Gaussian noise (AWGN) channel in which at each time $i$, the output of the channel $Y_i$ is the sum of the input $X_i$ and Gaussian noise $Z_i$.  Shannon showed in his original paper~\cite{Shannon48}  that launched the field of information theory that the capacity of the AWGN channel is 
\begin{equation}
\rvC(P) = \frac{1}{2}\log(1+P), \label{eqn:cap} 
\end{equation}
where $P$ is the signal-to-noise ratio (SNR). More precisely, let $M^*(W^n,\eps,P)$ be the maximum number of codewords that can be transmitted over $n$ independent uses of an AWGN channel with  SNR  $P$ and average error probability not exceeding $\eps\in (0,1)$. Then,  combining the direct part in \cite{Shannon48} and the strong converse by Shannon in~\cite{Sha59b} (also see Yoshihara~\cite{Yoshihara} and Wolfowitz~\cite{Wolfowitz}), one sees that   
\begin{equation}
\lim_{n\to\infty}\frac{1}{n}\log M^*(W^n,\eps,P)=\rvC(P)  \quad \mbox{bits per channel use} 
\end{equation}
holds for every $\eps\in (0,1)$. 

Recently, there has been  significant  renewed interest in studying the higher-order terms in the asymptotic expansion  of   non-asymptotic fundamental limits such as $\log M^*(W^n,\eps,P)$. This line of analysis was pioneered by Strassen \cite[Theorem~1.2]{Strassen}  for discrete memoryless channels (DMCs) and  is useful because it provides key insights into the amount of backoff from channel capacity for  block codes of finite  length $n$. For the AWGN channel, Hayashi \cite[Theorem~5]{Hayashi09} showed that 
\begin{equation}
\log M^*(W^n,\eps,P) = n\rvC(P) + \sqrt{n\rvV(P)}\Phi^{-1}(\eps) + o(\sqrt{n}) \label{eqn:hayashi}
\end{equation}
where $\Phi^{-1}(\cdot)$ is the inverse of the Gaussian cumulative distribution function and 
\begin{equation}
\rvV(P) = \log^2 \rme\cdot\frac{P(P+2)}{2(P+1)^2}    \quad \mbox{bits$^2$ per channel use}  \label{eqn:disp}
\end{equation}
is termed the {\em Gaussian dispersion function}~\cite{PPV10}. The  first two terms in the expansion in \eqref{eqn:hayashi} are collectively  known the {\em normal approximation}.  The functional form of $\rvV(P)$ was already known to Shannon~\cite[Section~X]{Sha59b} who analyzed the behavior of the reliability function of the AWGN channel at rates close to capacity. Subsequently, the $o(\sqrt{n})$ remainder term in the expansion in~\eqref{eqn:hayashi} was refined by Polyanskiy-Poor-Verd\'u~\cite[Theorem~54, Eq.~(294)]{PPV10} who showed that 
\begin{equation}
O(1)\le\log M^*(W^n,\eps,P) - \Big( n\rvC(P) + \sqrt{n\rvV(P)}\Phi^{-1}(\eps)  \Big) \le\frac{1}{2}\log n + O(1). \label{eqn:ppv_gauss}
\end{equation} 
The same bounds hold  under the maximum probability of error formalism. 

Despite these impressive advances in the fundamental limits of coding over a Gaussian channel, the gap in the third-order term beyond the normal approximation in \eqref{eqn:ppv_gauss} calls for further investigations. The authors of the present paper showed for  DMCs  with positive $\eps$-dispersion that the third-order term is no larger than $\frac{1}{2}\log n + O(1)$ \cite[Theorem~1]{TomTan12}, matching a lower bound by Polyanskiy~\cite[Theorem~53]{Pol10} for non-singular channels (also called channels with positive reverse dispersion~\cite[Eq.~(3.296)]{Pol10}). Altu\u{g} and Wagner~\cite{altug13} showed for singular, symmetric DMCs that the third-order term is $O(1)$. Moulin~\cite{mou13b} recently showed for a large class of channels (but {\em not} the AWGN channel) that the third-order term is  $\frac{1}{2}\log n + O(1)$. In light of these existing results for DMCs, a reasonable conjecture would be that the third-order term for the Gaussian case is either  $O(1)$ or $\frac{1}{2}\log n + O(1)$. In this paper, we show that in fact, the lower bound in \eqref{eqn:ppv_gauss} is loose. In particular, we establish that it can be improved to match the upper bound $\frac{1}{2}\log n + O(1)$.   Our proof technique is similar to that developed by Polyanskiy~\cite[Theorem 53]{Pol10} to show that $\frac{1}{2}\log n + O(1)$ is achievable for non-singular DMCs. However,  our proof is more involved due to the presence of power constraints on the codewords. 


\section{Problem Setup and Definitions}
Let $W$ be an AWGN channel where the noise variance\footnote{The assumption that the noise variance is $1$ does not entail any loss of generality because we can simply scale the admissible power   accordingly to ensure that the SNR is $P$.} is $1$, i.e.
\begin{equation}
W(y|x)=\frac{1}{\sqrt{2\pi}}\exp\Big(-\frac{(y-x)^2}{2} \Big).
\end{equation}
Let $\bx=(x_1,\ldots,x_n)$ and $\by=(y_1,\ldots, y_n)$ be two vectors in $\bbR^n$. Let $W^n(\by|\bx)=\prod_{i=1}^n W(y_i|x_i)$ be the $n$-fold   memoryless extension of $W$. An {\em $(n,M,\eps,P)_{\mathrm{av}}$-code} for the AWGN channel  $W$ is a system $\{ (\bx(m), \calD_m)\}_{m=1}^M$ where $\bx(m)\in\bbR^n,m \in\{ 1,\ldots, M\}$, are the codewords satisfying the maximal power constraint $\|\bx(m)\|_2^2\le n P$, the sets $\calD_m\subset\bbR^n$ are disjoint decoding regions  and the {\em average probability of error} does not exceed $\eps$, i.e.\
\begin{equation}
\frac{1}{M}\sum_{m=1}^M W^n\big( \calD_m^c \,\big|\, \bx(m)\big)\le \eps.
\end{equation}
Define $M^*(W^n,\eps,P) :=\max\big\{ M \in\bbN : \exists \, \mbox{ an } (n,M,\eps,P)_{\mathrm{av}}\mbox{-code for } W\big\}$.  

We also employ the Gaussian cumulative distribution function 
\begin{equation}
\Phi(a) := \int_{-\infty}^a \frac{1}{\sqrt{2\pi}}\exp\Big( -\frac{u^2}{2}\Big)\,\rmd u
\end{equation}
and define its inverse as $\Phi^{-1}(\eps): =\sup\{a\in\bbR: \Phi(a)\le \eps\}$, which evaluates to the usual inverse for $0 <\eps < 1$ and continuously extends to take values $\pm\infty$ outside that range.

\section{Main Result and Remarks}
Let us reiterate our main result.
\begin{theorem} \label{thm:ach3}
For all $0<\eps<1$  and $P \in (0,\infty)$, 
\begin{equation}
\log M^*(W^n,\eps,P) \ge  n \rvC(P)+\sqrt{n\rvV(P)}\Phi^{-1} (\eps) + \frac{1}{2}\log n + O(1) \label{eqn:sizeM_star}
\end{equation}
where  $\rvC(P)$ and $\rvV(P)$ are   the Gaussian capacity  and dispersion functions respectively.
\end{theorem}

We make the following remarks before proving the theorem in the following section.
\begin{enumerate}
\item As mentioned in  the Introduction, the upper bound on $\log M^*(W^n,\eps,P)$ in \eqref{eqn:ppv_gauss} was first established  by Polyanskiy-Poor-Verd\'u~\cite[Theorem~65]{PPV10}. They evaluated the meta-converse~\cite[Theorem~28]{PPV10} and appealed to the spherical symmetry in the Gaussian problem. The third-order term  in the normal approximation was shown to be upper bounded by $\frac{1}{2}\log n+O(1)$ (under the average or maximum error probability formalism). Thus, one has
\begin{equation}
\log M^*(W^n,\eps,P) =  n \rvC(P)+\sqrt{n\rvV(P)}\Phi^{-1} (\eps) + \frac{1}{2}\log n + O(1) . \label{eqn:sizeM_eq}
\end{equation}
The technique developed by the present authors in \cite{TomTan12} can also be used to prove the $\frac{1}{2}\log n + O(1)$ upper bound on the third-order term.

\item Our     strategy  for proving \eqref{eqn:sizeM_star} parallels that for non-singular DMCs  without cost constraints by Polyanskiy~\cite[Theorem~53]{Pol10}. It leverages on the random-coding union (RCU) bound~\cite[Theorem~16]{PPV10} and uses the log-likelihood ratio as the decoding metric, i.e.\ we do maximum likelihood decoding.  However,  the Gaussian problem involves cost (power) constraints and our random codebook generation strategy (which is similar to Shannon's~\cite{Sha59b}) involves drawing codewords independently and uniformly at random from the power sphere. Thus, a  more delicate analysis (vis-\`a-vis~\cite[Theorem~53]{Pol10}) is required. In particular, one cannot directly employ the  refined large-deviations  result stated in \cite[Lemma~47]{PPV10} which is crucial in showing the achievability of $\frac{1}{2}\log n+O(1)$.  This is because  \cite[Lemma~47]{PPV10} requires independence of a collection random variables whereas the independence structure is lacking in the AWGN problem.


\item In Theorem~\ref{thm:ach3}, we considered a maximal power constraint on the codewords, i.e.\ $\|\bx(m)\|_2^2\le nP$ for all $m$. It is easy to show that the third-order term is the same for the case of equal power constraints, i.e.\ $\|\bx(m)\|_2^2= nP$ for all $m$.   However, the strong converse does not even hold~\cite[Theorem~77]{Pol10} under the {\em average probability of error} formalism and the {\em average power constraint across the codebook}, i.e.\  $\frac{1}{M}\sum_{m=1}^M\|\bx(m)\|_2^2\le nP$. The $\eps$-capacity depends on $\eps$. We do not consider this case in this paper. Nonetheless, the strong converse and normal approximation do  hold~\cite[Theorem~54]{PPV10} under the {\em maximum  probability of error} formalism and average power constraint across the codebook but we do not consider this setup here. It is known~\cite[Eq.~(295)]{PPV10} that the third-order term is sandwiched between $O(1)$ and $\frac{3}{2}\log n + O(1)$.

\item A straightforward extension  of our proof technique (in particular, the application of Lemma~\ref{lem:boundU} in Section~\ref{sec:interval}) shows that the   achievability of $\frac{1}{2}\log n+O(1)$ also holds for the problem of information transmission over {\em parallel Gaussian channels} \cite[Section~9.4]{Cov06} in which the capacity is given by the well-known {\em water-filling} solution. See Appendix \ref{app:parallel} for a description of the modifications to  the proof of Theorem~\ref{thm:ach3} to this setting. This improves on the result in \cite[Theorem~81]{Pol10} by $\frac{1}{2}\log n$. However, this third-order achievability result does not match the converse bound given in  \cite[Theorem~80]{Pol10} in which it is shown that the third-order term is upper bounded by $\frac{k+1}{2}\log n + O(1)$ where $k\ge 1$ is the number of parallel Gaussian channels. We leave the closing of this gap for future research.

\item Finally, we make an observation concerning the relation between prefactors in the error exponents regime and the third-order terms in the normal approximation. In \cite{Sha59b}, Shannon  derived exponential bounds on the average error probability of optimal codes over a Gaussian channel using geometric arguments. For {\em high rates} (i.e.\ rates above the critical rate and below capacity), he showed that \cite[Eqs.~(4)--(5)]{Sha59b} 
\begin{equation}
\rmP_{\rme}^*(M,n)=\Theta\Big( \frac{\exp(-n F(\varphi))}{\sqrt{n}}\Big) \label{eqn:shannon_exponent}
\end{equation}
where $\rmP_{\rme}^*(M,n)$ is the    optimal average probability of error of a length-$n$ block code of size $M \in\bbN$, $\varphi=\varphi(R)$ is a cone angle related to the  signaling rate $R :=\frac{1}{n}\log M$  as follows \cite[Eq.~(28)]{Sha59b} 
\begin{align}
\exp(-nR ) =  \frac{\big(1+O\big(\frac{1}{n}\big)\big)\sin^n\varphi}{\sqrt{2\pi n} \, \sin\varphi \, \cos\varphi}  , \label{eqn:theta1_Rn}
\end{align}
and the exponent in \eqref{eqn:shannon_exponent} is defined as 
\begin{align}
F(\varphi)&:=\frac{P }{2}-\frac{\sqrt{P} \,G \, \cos\varphi}{2}-\log\big(G\sin\varphi\big),  \quad\mbox{where}\\
 G=G(\varphi ) & := \frac{1}{2}\big(\sqrt{P} \cos\varphi+ \sqrt{P\cos^2 \varphi+4}\big).
\end{align}
Furthermore for high rates, the error exponent (reliability function) of an AWGN channel is known and equals the sphere-packing exponent~\cite[Eq.~(7.4.33)]{gallagerIT} 
\begin{equation}
E(R) = \frac{P}{4\beta}\bigg( (\beta+1) -(\beta-1)\sqrt{ 1+\frac{4\beta}{P(\beta-1)}} \bigg)+\frac{1}{2}\log\bigg( \beta-\frac{P(\beta-1)}{2}\bigg[\sqrt{ 1+\frac{4\beta}{P(\beta-1)}}-1\bigg]\bigg) \label{eqn:ER}
\end{equation}
where $\beta :=\exp(2R)$. Simple algebra shows that $F(\theta )= E(\tilR(\theta))$ when $\tilR(\theta):= -\log\sin\theta$. Thus,
\begin{align}
F\big(\varphi(R) \big) &= E\big( \tilR(\varphi(R) )\big)   \\
&= E\big(-\log\sin (\varphi(R)) \big) \\
& = E\Big(R-\frac{\log n}{2n}+ \Theta\Big(\frac{1}{n}\Big)\Big)   \label{eqn:use_theta1_Rn}\\
&= E(R) - E'(R) \frac{\log n}{2n}+\Theta\Big(\frac{1}{n}\Big), \label{eqn:taylor_ER}
\end{align}
where \eqref{eqn:use_theta1_Rn} follows from \eqref{eqn:theta1_Rn} and \eqref{eqn:taylor_ER} follows by Taylor expanding the continuously differentiable function $E(R)$.  Note that $E'(R)\le 0$.  This leads to the conclusion that for high rates,
\begin{equation}
\rmP_{\rme}^*(M,n)=\Theta\Big( \frac{\exp(-nE(R))}{n^{(1+ | E'(R)| )/2}}\Big).
\end{equation}
Thus, the prefactor  of the AWGN channel is  $\Theta(n^{-(1+ | E'(R) | )/2})$. We showed in Theorem~\ref{thm:ach3} that the third-order term is  $\frac{1}{2}\log n+O(1)$. Somewhat surprisingly, this is analogous to the symmetric, discrete memoryless case. Indeed for non-singular, symmetric  DMCs (such as the binary symmetric channel) the prefactor in the error exponents regime  for high rates is $\Theta(n^{-(1+ |E'(R) |)/2})$ \cite{altug11,altug12,   altug12a, Sca13}    and for DMCs with positive $\eps$-dispersion, the third-order term is $\frac{1}{2}\log n+O(1)$ (combining \cite[Theorem~1]{TomTan12} and \cite[Theorem~53]{Pol10}). (Actually symmetry is not required for the third-order term to be $\frac{1}{2}\log n+O(1)$.) On the other hand,  for  singular, symmetric  DMCs (such as the binary erasure channel), the prefactor   is $\Theta(n^{-1/2})$ \cite{altug12,altug11, altug12a, Sca13}  and the third-order term is $O(1)$ (combining \cite[Proposition~1]{altug13} and \cite[Theorem~45]{PPV10}).  Also see~\cite[Theorem~23]{Pol13}. These results suggest  a  connection between prefactors and third-order terms. Indeed, a precise understanding of this connection  is a promising avenue for further research.

\end{enumerate}

\section{Proof of Theorem~\ref{thm:ach3}}
The proof, which is based on random coding, is split into several steps. 

\subsection{Random Codebook Generation And Encoding}
We first start by defining  the random coding distribution 
\begin{equation}
f_{\bX} (\bx) := \frac{\delta ( \|\bx\|_2^2   -nP )}{S_n(\sqrt{nP})} \label{eqn:rc_dist}
\end{equation}
where $\delta(\cdot)$ is the Dirac delta and $S_n(r) = \frac{2\pi^{n/2}}{\Gamma(n/2)}r^{n-1}$ is the surface area of a radius-$r$ sphere  in $\bbR^n$. We sample  $M$ length-$n$ codewords independently from $f_{\bX}$. In other words, we draw codewords uniformly at random from the surface   of the sphere in $\bbR^n$ with  radius $\sqrt{nP}$.  The number of codewords $M$ will be specified at the end of the proof in \eqref{eqn:logM}. These codewords are denoted as $\bx(m) = (x_1(m),\ldots, x_n(m)), m \in \{ 1,\ldots, M\}$.  To send message $m$, transmit codeword~$\bx(m)$. 

\subsection{Maximum-Likelihood Decoding }
Let the induced output density be $f_{\bX}W^n$, i.e.\
\begin{equation}
f_{\bX}W^n(\by):=\int_{\bx'} f_{\bX}(\bx')W^n(\by|\bx')\,\rmd\bx'  .
\end{equation}
Given $\by=(y_1,\ldots, y_n)$, the decoder selects the   message $m$ satisfying
\begin{equation}
 q(\bx(m),\by)> \max_{ \tilm \in \{1,\ldots, M\}\setminus \{m\}} q(\bx(\tilm),\by) , \label{eqn:decode-rule}
\end{equation}
where the decoding metric is the log-likelihood ratio defined as 
\begin{equation}
q(\bx,\by) := \log \frac{W^n(\by|\bx)}{f_{\bX}W^n(\by)}  . \label{eqn:decoding_metric}
\end{equation}  
If there is no unique $m\in\{1,\ldots, M\}$ satisfying \eqref{eqn:decode-rule}, declare an error. (This happens with probability zero.)

Since the denominator in \eqref{eqn:decoding_metric}, namely  $f_{\bX}W^n(\by)$,  is constant across all codewords, this is simply maximum-likelihood or, in this Gaussian case, minimum-Euclidean distance decoding.
We will take advantage of the latter observation in our proof, more precisely the fact that
\begin{equation}
  q(\bx,\by) = \frac{n}{2} \log \frac{1}{2\pi} + \langle \bx, \by \rangle - nP - \|\by\|_2^2 - \log f_{\bX}W^n(\by) \label{eqn:inner_product}
\end{equation}
only depends on the codeword through the inner product $\langle \bx, \by \rangle=\sum_{i=1}^n x_i y_i$. In fact, $q(\bx,\by)$ is equal to $\langle \bx, \by \rangle$ up to a shift that only depends on $\|\by\|_2^2$.

Note that because $f_{\bX}W^n$ is not a product density, $q(\bx,\by)$ is {\em not separable} (into a sum of $n$ terms) unlike in the   i.i.d.\ random coding  case~\cite[Theorem~53]{Pol10}.

\subsection{The Random Coding Union (RCU) Bound}
All the randomly drawn codewords satisfy the cost constraints with probability one. By using the same proof technique as that for the RCU bound~\cite[Theorem~16]{PPV10}, we may assert that  there exists an $(n,M,\eps', P)_{\mathrm{av}}$-code satisfying 
\begin{equation}
\eps'\le \bbE\left[ \min\big\{1,M \Pr \big( q(\bar{\bX} , \bY )  \ge q(\bX , \bY )  |\bX,\bY \big)\big\}\right] \label{eqn:rcu}
\end{equation}
where the random variables $(\bar{\bX},\bX,\bY)$ are distributed as $f_{\bX}(\bar{\bx})\times f_{\bX}(\bx)\times W^n(\by|\bx)$.  Now, introduce the function 
\begin{equation}
g(t,\by) := \Pr\big(q(\bar{\bX} , \bY )  \ge t  \,\big|\, \bY=\by \big). \label{eqn:gty}
\end{equation}
Since $\bar{\bX}$ is independent of $\bX$,  the   probability in \eqref{eqn:rcu} can be written as 
\begin{equation}
 \Pr \big( q(\bar{\bX} , \bY )  \ge q(\bX , \bY )  |\bX,\bY \big) = g(q(\bX , \bY ) ,\bY).
\end{equation}
Furthermore, by Bayes rule, we have  $f_{\bX|\bY}(\bx|\by)\times f_{\bX}W^n(\by)= f_{\bX}(\bx) \times W^n(\by|\bx)$ and so
\begin{equation}
f_{\bX}(\bar{\bx})=f_{\bX}(\bar{\bx})\frac{f_{\bX|\bY}(\bar{\bx}|\by)}{f_{\bX|\bY}(\bar{\bx}|\by)} = f_{\bX|\bY}(\bar{\bx}|\by)\exp(-q(\bar{\bx} , \by)).
\end{equation}
For a fixed sequence $\by \in \bbR^n$ and a constant $t\in\bbR$, multiplying both sides by $\bone\{q (\bar{\bx} , \by)\ge t\}$ and integrating over all $\bar{\bx}$ yields  the following alternative representation of $g(t,\by)$: 
\begin{equation}
g(t,\by) =\bbE\big[  \exp(-q( \bX,\bY))\bone\{ q( \bX,\bY)\ge t  \} \,\big|\, \bY=\by\big]. \label{eqn:integrating}
\end{equation}

\subsection{A High-Probability Set}
Consider the set of  ``typical'' channel outputs whose norms are approximately $\sqrt{n(P+1)}$. More precisely, define
\begin{equation}
\calF:=\Big\{ \by\in\bbR^n : \frac{1}{n} \|\by\|_2^2 \in [ P+1 -\delta ,  P+1+\delta]\Big\}.
\end{equation}
We claim that the probability of $\bY\in\calF$ is large.  First   the union bound yields
\begin{align}
\Pr(\bY\in\calF^c)  \le  \Pr\bigg( \frac{1}{n}\|\bX+\bZ\|_2^2 > P+1+\delta\bigg)  + \Pr\bigg( \frac{1}{n}\|\bX+\bZ\|_2^2 < P+1 - \delta\bigg) .
\end{align}
Since the bounding of both probabilities can be done in a similar fashion, we focus on the first which may be written as
\begin{equation}
\Pr\bigg( \frac{1}{n}\|\bX+\bZ\|_2^2 > P+1+\delta\bigg)= \Pr\bigg( \frac{1}{n}\big(2\langle\bX,\bZ\rangle + \|\bZ\|_2^2\big) > 1+\delta\bigg).\label{eqn:dropP}
\end{equation}
Define the following ``typical'' set   of noises
\begin{equation}
\calG:=\Big\{ \bz\in\bbR^n:\frac{1}{n} \|\bz\|_2^2 \le 1+\frac{\delta}{2} \Big\}  .
\end{equation}
Since $\bZ=(Z_1,\ldots, Z_n)\sim\calN(\bzero,\bI_{n\times n})$, by the Chernoff bound (or, more precisely, by Cramer's theorem~\cite[Theorem~2.2.3]{Dembo} for $\chi_1^2$ random variables), the probability that $\bZ\in\calG^c$ is upper bounded by $\exp(-\kappa_1 n \delta^2)$ for some constant  $\kappa_1>0$.  Now, we continue bounding the probability in \eqref{eqn:dropP} as follows:
\begin{align}
\Pr\bigg( \frac{1}{n}\big(2\langle\bX,\bZ\rangle + \|\bZ\|_2^2\big) > 1+\delta\bigg) &\le\Pr\bigg( \frac{1}{n}\big(2\langle\bX,\bZ\rangle \!+\! \|\bZ\|_2^2\big) > 1+\delta\,\bigg|\, \bZ\in\calG\bigg)\Pr(\bZ\in\calG) \! +\!\Pr(\bZ\in\calG^c) \\
&\le \Pr\bigg( \frac{2}{n} \langle\bX,\bZ\rangle > \frac{\delta}{2}\,\bigg|\,\bZ\in \calG\bigg)\Pr(\bZ\in\calG) +\Pr(\bZ\in\calG^c)  \label{eqn:usedefG}\\
&\le \Pr\bigg( \frac{1}{n}\sum_{i=1}^n X_i Z_i> \frac{\delta}{4}\bigg) +\Pr(\bZ\in\calG^c) ,
\end{align}
where in~\eqref{eqn:usedefG} we used the definition of $\calG$. 
By spherical symmetry, we may take $\bX$ to be any point on the power sphere $\{\bx:\|\bx\|_2^2 = nP\}$. We take $\bX$ to be equal to $(\sqrt{nP}, 0, \ldots, 0)$. Then the first term reduces to 
\begin{equation}
\Pr\bigg(   Z_1 > \frac{\delta}{4} \cdot\sqrt{\frac{n}{P}} \, \bigg) =1-\Phi\bigg( \frac{\delta}{4} \cdot\sqrt{\frac{n}{P}} \, \bigg)\le\exp(-\kappa_2 n \delta^2) ,
\end{equation}
where $\kappa_2 >0$ is a  constant.  By putting all the bounds together and setting $\delta=n^{-1/3}$, we deduce that 
\begin{equation}
\Pr(\bY\in\calF)\ge 1-\xi_n\label{eqn:chernoff}
\end{equation}
where $\xi_n := \exp(-\kappa_3 n^{1/3})$ for some $\kappa_3>0$. Note that $\xi_n$ decays faster than any polynomial.

\subsection{Probability Of The Log-Likelihood Ratio Belonging To An Interval} \label{sec:interval}

We would like to upper bound $g(t,\by)$ in \eqref{eqn:gty} to evaluate the RCU bound. This we do in the next section. As an intermediate step,   we consider the problem of upper bounding 
\begin{equation}
h(\by; a, \mu) := \Pr\big( q(\bX,\bY)\in [a,a+\mu] \, \big| \, \bY = \by\big) ,
\end{equation}
where $a \in \mathbb{R}$ and $\mu > 0$ are some constants. Because $\bY$ is fixed to some constant vector $\by$ and $\|\bX\|_2^2$ is also constant, $h(\by; a, \mu)$ can be rewritten using~\eqref{eqn:inner_product} as
\begin{equation}
h(\by;a,\mu) := \Pr\big( \langle\bX,\bY\rangle\in [a', a' + \mu] \, \big| \, \bY = \by\big) ,  \label{eqn:inner_prod}
\end{equation}
for some other constant $a' \in \mathbb{R}$. It is clear that $h(\by; a, \mu)$ depends on $\by$ through its norm and so we may define (with an abuse of notation),
\begin{equation}
h(s;a,\mu) : = h(\by;a,\mu) ,\quad\mbox{if}\quad s = \frac{1}{n}\|\by\|_2^2.
\end{equation}
In the rest of this section, we assume that $\by\in\calF$ or, equivalently, $s\in [ P+1 -\delta ,  P+1+\delta]$. 

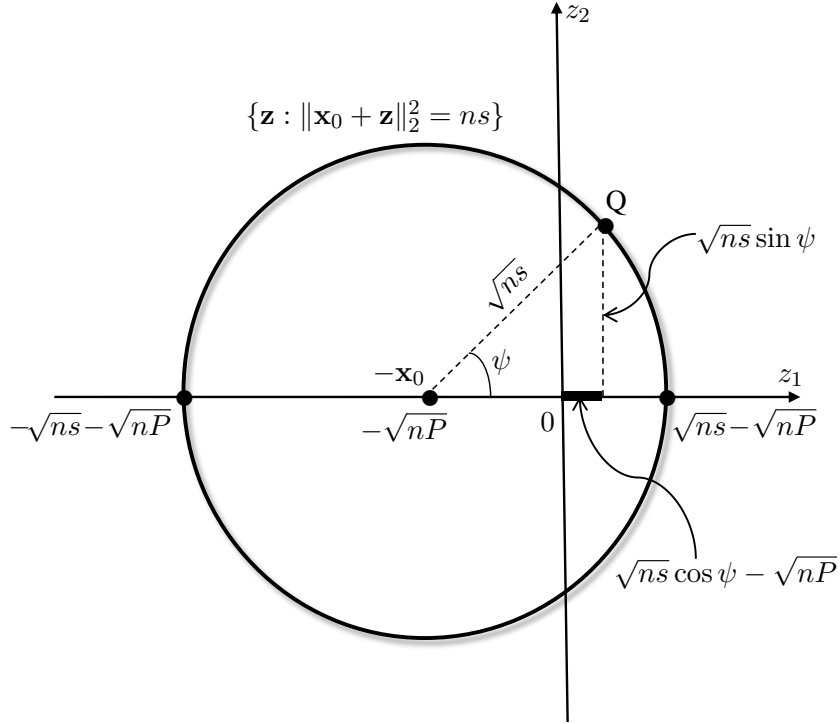
\begin{figure}[t]
\centering
\begin{overpic}[width=.6\columnwidth]{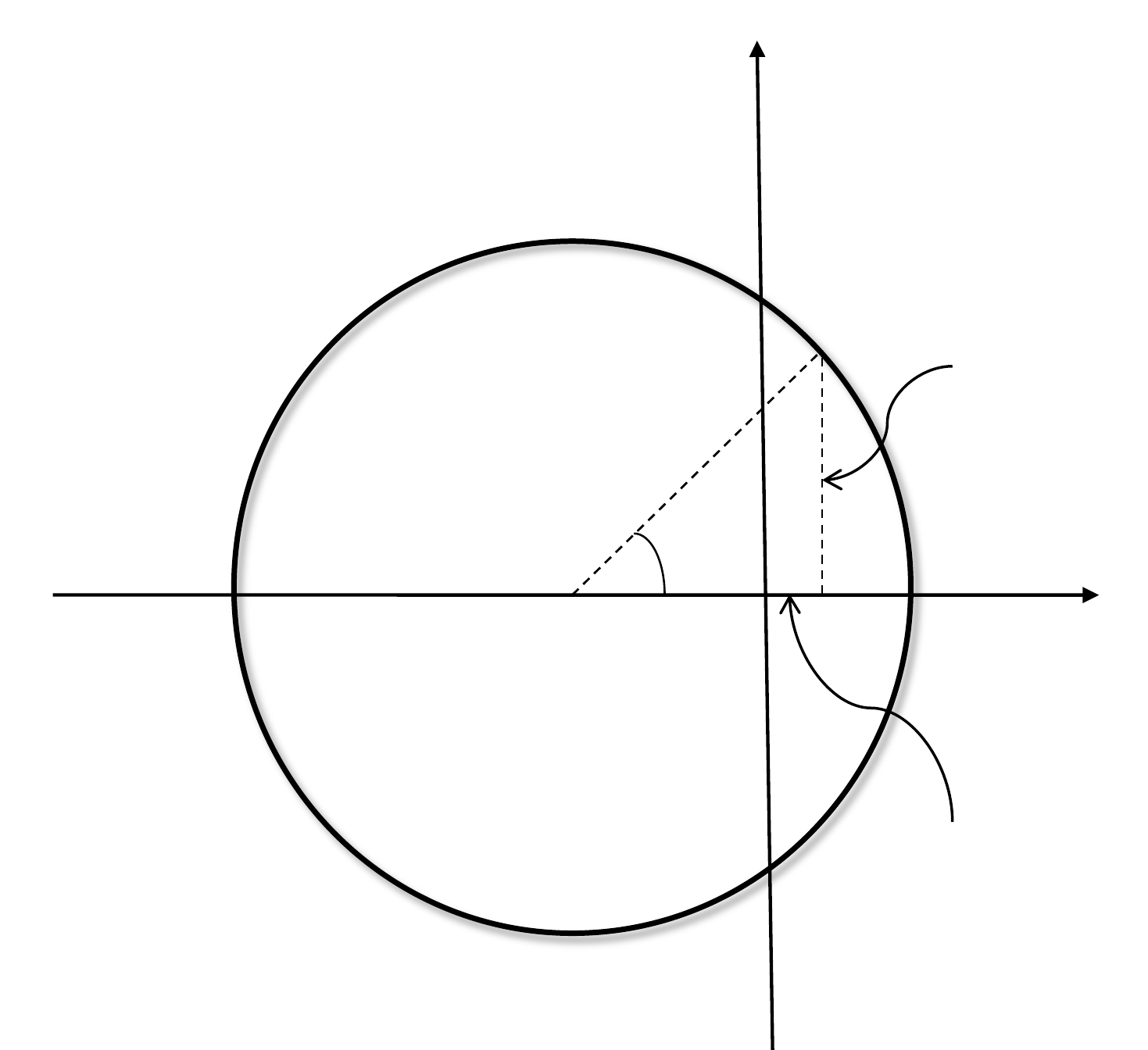}
\put(93,41.5){$z_1$}
\put(67,86){$z_2$}
\put(64,35.5){$0$}
\put(42,34.5){$-\sqrt{nP}$}
\put(43.5,41.6){$-\bx_0$}
\put(72,62.5){Q}
\put(80,35){$\sqrt{ns}\!-\!\sqrt{nP}$}
\put(-1,35){$-\!\sqrt{ns}\!-\!\sqrt{nP}$}
\put(56,50){\rotatebox{45}{$\sqrt{ns}$}}
\put(83,58){$\sqrt{ns}\sin\psi$}
\put(73,17){$\sqrt{ns}\cos\psi-\sqrt{nP}$}
\put(58,43){$\psi$}
\put(79.5,39.5){\circle*{2}}
\put(20.5,39.5){\circle*{2}}
\put(50.5,39.5){\circle*{2}}
\put(72,60.5){\circle*{2}}
\put(28,73){$\{\bz:\|\bx_0+\bz\|_2^2=ns\}$}
\linethickness{1.2mm}
\put(66.5,39.8){\line(1,0){5}}
\end{overpic}
\caption{Illustration of the relation between $Z_1$ and $\Psi$ in \eqref{eqn:ZTheta} in two dimensions.   The transformation of this figure to the $U$ coordinate system via~\eqref{eqn:UTheta} translates the sphere to the origin and scales its radius to be $1$.  }
\label{fig:z1}
\end{figure}    

By introducing the standard Gaussian random vector $\bZ=(Z_1, \ldots, Z_n)\sim \calN(\bzero,\bI_{n\times n})$, we have 
\begin{align}
h(s;a,\mu)
&= \Pr\left( \langle\bX,\bX+\bZ\rangle\in [a',a'+\mu] \, \Big|\, \|\bX+\bZ\|_2^2 = ns\right) \\
&= \Pr\bigg(   \sum_{i=1}^n X_i Z_i + nP \in [a',a'+\mu] \, \bigg|\, \|\bX+\bZ\|_2^2 = ns \bigg)   \label{eqn:a_prime}
\end{align}
where~\eqref{eqn:a_prime} follows by  the observation that $\langle\bX,\bX\rangle = nP$ with probability one. Now, define 
\begin{equation}
\bx_0  := \big( \sqrt{nP}, 0,\ldots, 0\big)
\end{equation}
to be a fixed vector on the power sphere. By spherical symmetry, we may pick $\bX$ in \eqref{eqn:a_prime} to be equal to $\bx_0$. Thus, we have 
\begin{equation}
h(s;a,\mu) = 
\Pr\bigg(    Z_1 + \sqrt{nP} \in \Big[ \frac{a'}{\sqrt{nP}},\frac{a'+\mu}{\sqrt{nP}}\Big] \, \bigg|\, \|\bx_0+\bZ\|_2^2 =ns \bigg)   . \label{eqn:x0}
\end{equation}
In other words, we are conditioning on the event that the random vector $\bZ \sim \calN(\bzero,\bI_{n\times n})$ lands on the surface of a sphere of radius $\sqrt{ns}$ centered at $-\bx_0 = (-\sqrt{nP}, 0, \ldots, 0)$.
See Fig.~\ref{fig:z1}.   We are then asking what is the probability that the first component 
plus $\sqrt{nP}$ belongs to the prescribed interval of length proportional to $\mu/\sqrt{n}$. 

Let us now derive the conditional density of $Z_1$ given the event  $\calE :=\{   \|\bx_0+\bZ\|_2^2 = ns\}$. Denote this density as $f_{Z_1|\calE} (z_1)$. Note that the support of $f_{Z_1|\calE}(z_1)$ is 
$[-\sqrt{ns}-\sqrt{nP},\sqrt{ns}-\sqrt{nP}]$.
It is easier to find the conditional density  of the angle $\Psi \in [0,2\pi]$ given the event $\calE$ where $\Psi$  and  $Z_1$ are related as follows:
\begin{equation}
Z_1 = \sqrt{ns} \cos\Psi - \sqrt{nP} .  
\label{eqn:ZTheta}
\end{equation}
Again see Fig.~\ref{fig:z1}.  Now, we have 
\begin{equation}
f_{\Psi|\calE}(\psi)\, \rmd \psi\propto \left(\sin^{n-2} \psi\right)  \exp\left( -\frac{n}{2} \left[ (\sqrt{s}\cos\psi - \sqrt{P}
)^2 + s\sin^2\psi\right]\right)\, \rmd \psi.
\end{equation}
This follows because  the area element (an $(n-1)$-dimensional annulus of radius $\sqrt{ns}\sin\psi$ and width $\rmd\psi$) is proportional to  $\sin^{n-2}\psi$ (similar to Shannon's derivation in~\cite[Eq.~(21)]{Sha59b}) and the Gaussian weighting is proportional to $\exp\big( -\frac{n}{2} \big[ (\sqrt{s}\cos\psi - \sqrt{P}
)^2 + s\sin^2\psi\big]\big)$.  This is just $\exp(-d^2 / 2)$ where $d$ is the distance of the point described by $\psi$ (point Q in Fig.~\ref{fig:z1}) to the origin. We are obviously leveraging heavily on the radial symmetry of the problem around the first axis. Now, we consider the change of variables  
\begin{equation}
U = \cos\Psi \label{eqn:UTheta}
\end{equation}
resulting in 
\begin{equation}
f_{U|\calE} (u) \, \rmd u \propto (1-u^2)^{(n-3)/2}\exp \big( n \sqrt{Ps} u \big) \, \rmd u.
\end{equation}
Note that $U$ takes   values in $[-1,1]$. More precisely,  the conditional density of $U$ given $\calE$ is 
\begin{equation}
f_{U|\calE} (u)  =\frac{1}{F_n}  (1-u^2)^{(n-3)/2} \exp\big(  n \sqrt{Ps} u \big)  \bone\{ u \in [-1,1]\} ,\label{eqn:densityU}
\end{equation}
where the normalization constant is
\begin{equation}
F_n := \int_{-1}^1 (1-u^2)^{(n-3)/2} \exp\big( n \sqrt{Ps} u \big)
\, \rmd u \label{eqn:Fn} .
\end{equation}
The conditional density we have derived in \eqref{eqn:densityU}--\eqref{eqn:Fn} reduces to that by Stam~\cite[Eq.~(3)]{Stam} for the limiting case $P=0$, i.e.\ the sphere is centered at the origin. 
It is of paramount importance to analyze how $\sup_{u \in [-1,1]} f_{U|\calE}(u)$ scales with $n$. The answer turns out to be $O(\sqrt{n})$. More formally, we state the following lemma whose proof is provided in  Appendix~\ref{app:boundU}.
\begin{lemma} \label{lem:boundU}
Define the function
\begin{equation}
L(P,s) := \frac{(2Ps)^2}{\sqrt{2\pi}}\cdot \sqrt{\frac{1+4Ps-\sqrt{1+4Ps}}{ ( \sqrt{1+4Ps}-1)^5}  } .\label{eqn:defL}
\end{equation}
The following bound holds:
\begin{align}
\limsup_{n\to\infty}\frac{1}{\sqrt{n}} \sup_{u\in [-1,1]} f_{U|\calE}(u) \le L(P,s). 
\end{align}
\end{lemma}

Equipped with this lemma, let us consider the probability $h(s;a,\mu)$ in \eqref{eqn:x0}. We have
\begin{align}
h(s;a,\mu) & = 
  \Pr\bigg(    \sqrt{ns} \, U   \in \Big[ \frac{a'}{\sqrt{nP}}, \frac{a' + \mu}{\sqrt{nP}} \Big]
  \,\bigg|\,\,  \calE\bigg) \label{eqn:ch_to_U} \\
&= 
\int_{ a'/ (n\sqrt{Ps})   }^{(a'+\mu)/ (n\sqrt{Ps} )   } f_{U|\calE}(u) \, \rmd u \label{eqn:aprime2}\\
&\le 
\int_{ a'/ (n\sqrt{Ps})   }^{(a'+\mu)/ (n\sqrt{Ps} )   }  2 \,   L(P,s) \, \sqrt{n} \, \rmd u  \label{eqn:use_lem}     \\
 &=    \frac{2\,   L(P,s)  \,  \mu}{\sqrt{n Ps}}, \label{eqn:final_bd}
\end{align}
where \eqref{eqn:ch_to_U} follows from the fact that $Z_1   = \sqrt{ns} \, U 
-\sqrt{nP}$ due to  \eqref{eqn:ZTheta} and \eqref{eqn:UTheta}, and~\eqref{eqn:use_lem} holds for all sufficiently large $n$ (depending only on $P$ and $s$) on account of Lemma~\ref{lem:boundU}.

Since $s\in  [ P+1 -\delta ,  P+1+\delta]$ and $\delta = n^{-1/3}\to 0$,  we deduce that for all  $\by\in\calF$  and $n$ sufficiently large (depending only on $P$),
\begin{equation}
h(\by;a,\mu) \le K(P)  \cdot\frac{ \mu }{\sqrt{n }} , \label{eqn:root_n}
\end{equation}
for some function $K(P)$. In fact, by the continuity of $s\mapsto L(P,s)$, the constant $K(P)$ can be taken to be
\begin{equation}
K(P)=\frac{3\, L(P,P+1)}{\sqrt{P (P+1)}}. 
\end{equation}

\subsection{Probability That The Decoding Metric Exceeds $t$ For An Incorrect Codeword}
We now return to bounding $g(t,\by)$ defined in \eqref{eqn:gty}. Again, we assume $\by\in\calF$. The idea here is to consider the   second form of $g(t,\by)$ in \eqref{eqn:integrating} and to slice the interval $[t,\infty)$ into non-overlapping segments $\{ [t+l\eta, t+(l+1)\eta): l \in \bbN\cup\{0\}\}$ where $\eta>0$ is a constant. Then we apply \eqref{eqn:root_n} to each segment. This is modelled after  the proof of \cite[Lemma~47]{PPV10}. Indeed, we have   
\begin{align}
g(t,\by)& =\bbE\big[  \exp(-q( \bX,\bY))\bone\{ q( \bX,\bY)\ge t  \} \,\big|\, \bY=\by\big]\nn\\
&\le \sum_{l=0}^{\infty} \exp(-t -l\eta) \Pr\left(t+l\eta\le q(\bX,\bY) < t+(l+1)\eta \,\big|\, \bY = \by\right) \label{eqn:slices} \\
&\le\sum_{l=0}^{\infty} \exp(-t -l\eta) \cdot \frac{ K(P)\, \eta}{ \sqrt{n}}  \label{eqn:use_previous}\\
&= \frac{\exp(-t )}{1-\exp(-\eta)}   \cdot\frac{K(P)\, \eta}{ \sqrt{n}}    . \label{eqn:geom}
\end{align}
Since $\eta$ is a free parameter, we may choose it to be $\log 2$ yielding
\begin{equation}
g(t,\by) \le \frac{G \, \exp(-t )}{\sqrt{n}}  \label{eqn:Lambda_bd}
\end{equation}
where $G=G(P)=(2\log 2)\, K(P)$.

\subsection{Evaluating The RCU Bound}
We now have all the necessary ingredients to evaluate the RCU bound in \eqref{eqn:rcu}. Consider,
 \begin{align}
\eps'&\le \bbE\left[ \min\big\{1,Mg(q(\bX,\bY) ,\bY) \big\}\right]\nn\\ 
&\le\Pr(\bY\in \calF^c)+  \bbE\left[ \min\big\{1,Mg(q(\bX,\bY) ,\bY) \big\}\,\Big|\, \bY\in\calF \right] \cdot \Pr(\bY \in \calF) \\
 & \le\Pr(\bY\in \calF^c) + \bbE\left[  \min\left\{1, \frac{M G  \exp(-q(\bX,\bY) ) }{\sqrt{n}} \right\} \,\bigg|\, \bY\in\calF \right] \cdot \Pr(\bY \in \calF)\label{eqn:use_integrating}\\
  & \le\xi_n  + \bbE\left[  \min\left\{1, \frac{MG\exp(-q(\bX,\bY) ) }{\sqrt{n}} \right\} \,\bigg|\, \bY\in\calF\right] \cdot \Pr(\bY \in \calF)\label{eqn:use_F_bound}
  \end{align}
where \eqref{eqn:use_integrating} is due to~\eqref{eqn:Lambda_bd} with $t = q(\bX,\bY)$  and \eqref{eqn:use_F_bound} uses the   bound in \eqref{eqn:chernoff}. Now we split the expectation into two parts depending on whether $q(\bx,\by)> \log (MG/\sqrt{n})$ or otherwise, i.e.\
\begin{align}  
& \bbE\left[  \min\left\{1, \frac{MG\exp(-q(\bX,\bY) ) }{\sqrt{n}} \right\} \,\bigg|\, \bY\in\calF\right] \nn\\
&\le \Pr\left( q(\bX,\bY) \le \log \frac{MG}{\sqrt{n}} \,\bigg|\,  \bY\in\calF\right) + \frac{MG}{\sqrt{n }} \bbE\left[ \bone\left\{ q(\bX,\bY)  > \log\frac{MG}{\sqrt{n}} \right\}\exp(-q(\bX,\bY)) \,\bigg|\, \bY\in\calF\right] \label{eqn:expand_min} .
\end{align}
By applying \eqref{eqn:Lambda_bd} with $t = \log (MG/\sqrt{n})$, we know that  the second term can be bounded as 
\begin{equation}
\frac{MG}{\sqrt{n }} \bbE\left[ \bone\left\{ q(\bX,\bY)  > \log\frac{MG}{\sqrt{n}} \right\}\exp(-q(\bX,\bY))  \,\bigg|\, \bY\in\calF\right]\le \frac{G}{\sqrt{n}}.
\end{equation}
Now let $f_{Y}^*(y) = \calN(y;0,P+1)$ be the capacity-achieving output distribution and $f_{\bY}^* (\by)=\prod_{i=1}^n f_{Y}^*(y_i)$ its $n$-fold memoryless extension. In Step 1 of the proof of Lemma~61 in~\cite{PPV10}, Polyanskiy-Poor-Verd\'u showed that  on $\calF$, the ratio of the  induced output density  $f_{\bX}W^n(\by)$ and  $f_{\bY}^*(\by)$   can be bounded by a finite  constant $J$, i.e.\
\begin{equation}
\sup_{\by\in\calF} \frac{  f_{\bX}W^n(\by)}{  f_{\bY}^*(\by)} \le J .\label{eqn:change_meas}
\end{equation}
Also see \cite[Proposition~2]{Mol13}. We return to bounding the first term in \eqref{eqn:expand_min}.  Using the definition of $q(\bx,\by)$ in~\eqref{eqn:decoding_metric} and applying the bound in~\eqref{eqn:change_meas}  yields
\begin{align}
\Pr\left( q(\bX,\bY) \le \log \frac{MG}{\sqrt{n}} \,\bigg|\,  \bY\in\calF \right) &=\Pr\left( \log\frac{W^n(\bY|\bX)}{f_{\bX}W^n(\bY)}  \le \log \frac{MG}{\sqrt{n}} \,\bigg|\,  \bY\in\calF \right) \\
& \le \Pr\left( \log\frac{W^n(\bY|\bX)}{f_{\bY}^*(\bY)} \le \log \frac{MGJ}{\sqrt{n}} \,\bigg|\,  \bY\in\calF\right)  . 
\end{align}
Thus, when we multiply the first term in \eqref{eqn:expand_min}   by $\Pr(\bY \in \calF)$, use Bayes rule and drop the event $\{\bY\in\calF\}$, we see that the product can be bounded as follows:
\begin{align}
\Pr\left( q(\bX,\bY) \le \log \frac{MG}{\sqrt{n}} \,\bigg|\,  \bY\in\calF \right) \cdot\Pr(\bY \in \calF)\le\Pr  \left( \log\frac{W^n(\bY|\bX)}{f_{\bY}^*(\bY)} \le \log \frac{MGJ}{\sqrt{n}} \right)\label{eqn:first_term} .
\end{align} 
The right-hand-side of \eqref{eqn:first_term} can be written as an average over $\bX\sim f_{\bX}$, i.e.\  
\begin{align}
\Pr  \left( \log\frac{W^n(\bY|\bX)}{f_{\bY}^*(\bY)} \le \log \frac{MGJ}{\sqrt{n}} \right)= \int_{\bx}f_{\bX}(\bx) \Pr\left( \log\frac{W^n(\bY|\bX)}{f_{\bY}^*(\bY)} \le \log \frac{M G J }{\sqrt{n}} \, \bigg|\, \bX=\bx \right)\,\rmd \bx   \label{eqn:pick_X}.
\end{align}
By noting that $f_{\bY}^*(\by)$ is a product density, 
\begin{equation}
 \Pr\bigg( \log\frac{W^n(\bY|\bX)}{f_{\bY}^*(\bY)} \le \log \frac{M G J }{\sqrt{n}} \, \bigg|\, \bX=\bx \bigg) =  \Pr\bigg(\sum_{i=1}^n \log\frac{W (Y_i| X_i)}{f_{Y}^*(Y_i)} \le \log \frac{M G J }{\sqrt{n}} \, \bigg|\, \bX=\bx \bigg).
\end{equation}
The above probability does not depend on $\bx$ as long as it is on the power sphere $\{\bx:\|\bx\|_2^2=nP\}$ because of  spherical symmetry. Hence we may take $\bx=(\sqrt{P},\ldots, \sqrt{P})$. It is then easy to check that  the first two central moments  of the information density are
\begin{align}
\bbE\left[\frac{1}{n}\sum_{i=1}^n \log\frac{W (Y_i| \sqrt{P})}{f_{Y}^*(Y_i)} \right]  = \rvC(P) , \quad\mbox{and}\quad \var\left[ \frac{1}{n}\sum_{i=1}^n \log\frac{W (Y_i| \sqrt{P})}{f_{Y}^*(Y_i)}\right]  = \frac{   \rvV(P) }{n}.
\end{align}
Furthermore, the following third-absolute moment 
\begin{equation}
\rvT(P) := \frac{1}{n} \sum_{i=1}^n  \bbE \left[\left| \log\frac{W (Y_i| \sqrt{P})}{f_{Y}^*(Y_i)}- \bbE\bigg[\log\frac{W (Y_i| \sqrt{P})}{f_{Y}^*(Y_i)}\bigg]\right|^3\right]
\end{equation} 
is  obviously bounded (note the scaling). See \cite[Lemma~10 and Appendix~A]{ScarlettTan} for a precise analysis of third absolute moments of information densities involving Gaussians.  This allows us to  apply  the Berry-Esseen theorem~\cite[Theorem~2 in Section~XVI.5]{feller}, which implies that 
\begin{equation}
\Pr\left( \log\frac{W^n(\bY|\bX)}{f_{\bY}^*(\bY)} \le \log \frac{M G J }{\sqrt{n}} \, \bigg|\, \bX= (\sqrt{P},\ldots, \sqrt{P})\right)\le\Phi\left( \frac{\log \frac{M G J }{\sqrt{n}} -n\rvC(P)}{\sqrt{n\rvV(P)}} \right) + \frac{6\, \rvT(P)}{\sqrt{n\rvV(P)^3}} .\label{eqn:be}  
\end{equation}
Let $B=B(P) :=  6\,\rvT(P) / \rvV(P)^{3/2}$. 
We deduce   that 
\begin{equation}
\Pr\left( \log\frac{W^n(\bY|\bX)}{f_{\bY}^*(\bY)} \le \log \frac{MGJ}{\sqrt{n}} \right)\le \Phi\left( \frac{\log \frac{M G J }{\sqrt{n}} -n\rvC(P)}{\sqrt{n\rvV(P)}} \right) + \frac{B}{\sqrt{n}} .
\end{equation}
 Putting all the bounds together, we obtain
\begin{equation}
\eps'\le \Phi\left( \frac{\log \frac{M G J }{\sqrt{n}} -n\rvC(P)}{\sqrt{n\rvV(P)}} \right) + \frac{B}{\sqrt{n}} + \frac{G}{\sqrt{n}}+\xi_n.
\end{equation}
Now choose
\begin{align}
\log M = n\rvC(P) + \sqrt{n \rvV(P)} \Phi^{-1}\left( \eps - \frac{B+G}{\sqrt{n}}-\xi_n \right) + \frac{1}{2}\log n-\log(GJ) \label{eqn:logM}
\end{align}
ensuring that 
\begin{equation}
\eps'\le \eps.
\end{equation}
Hence, there exists an  $(n,M,\eps,P)_{\mathrm{av}}$-code  where $M$ is given by \eqref{eqn:logM}.  It is easily seen by Taylor expanding $\Phi^{-1}(\cdot)$ around $\eps$ that 
\begin{equation}
\log M =  n\rvC(P) + \sqrt{n \rvV(P)} \Phi^{-1}(\eps) + \frac{1}{2}\log n+O(1).
\end{equation}
This completes the proof of Theorem~\ref{thm:ach3}.\qed

\appendices
\numberwithin{equation}{section}
\section{Modifications of the Proof to the    Parallel Gaussian Channels Settng} \label{app:parallel}
In this appendix, we give a sketch of how the proof of Theorem~\ref{thm:ach3} can be used for the scenario where information is to be transmitted across $k$ parallel Gaussian channels. See Section 9.4 of \cite{Cov06} for the precise problem setting. Let the input and output to the channel be $(\bX_1, \ldots, \bX_k)$ and $(\bY_1, \ldots, \bY_k)$ respectively. Let the  independent noises of each of the channels have variances $N_1,\ldots, N_k$ and denote the total admissible power as $P$. Let $|\cdot|^+:=\max\{0,\cdot\}$ and set $P_1,\ldots, P_k$ be the power assignments  that maximize the information capacity expression, i.e.\
\begin{equation}
P_j = |\nu-N_j|^+ \label{eqn:water}
\end{equation}
where the Karush-Kuhn-Tucker multiplier $\nu$ is chosen to satisfy the total power constraint
\begin{equation}
\sum_{j=1}^k |\nu-N_j|^+=P. \label{eqn:kkt}
\end{equation}
Let $\calP^+:=\{j  \in\{1,\ldots, k\}: P_j>0\}$. Clearly, \eqref{eqn:water} and \eqref{eqn:kkt} imply that  $\calP^+$ is non-empty if $P>0$.
We use the random coding distribution $f_{\bX_1}\times\ldots\times f_{\bX_k}$ where each constituent distribution $f_{\bX_j}$ is given by~\eqref{eqn:rc_dist} with $P_j$ in place of $P$ there. Close inspection of the proof of Theorem~\ref{thm:ach3} shows that the only estimate that needs to be verified is~\eqref{eqn:final_bd}. For this, we consider the analogue of \eqref{eqn:a_prime} which can be written as
\begin{equation}
h(s_1,\ldots, s_k;a,\mu)=\Pr\bigg(\sum_{j=1}^k  \sqrt{P_j}\, Z_{j1}\in \Big[\frac{a_2}{\sqrt{n}}, \frac{a_2+\mu}{\sqrt{n}}\Big] \,\bigg|\, \|\bX_j+\bZ_j\|_2^2 = ns_j,\,\forall\, j  \in\{1,\ldots, k\}   \bigg)   , \label{eqn:Zk}
\end{equation}
where $a_2$ is related to $a'$ in \eqref{eqn:a_prime} by a constant shift. Note that the sum of the inner products $\sum_{j=1}^k\langle\bX_j,\bY_j\rangle$ in the analogue of \eqref{eqn:inner_prod} reduces to $\sum_{j=1}^k\sqrt{P_j} Z_{j1}= \sum_{j\in \calP^+}\sqrt{P_j}Z_{j1}$ once we have exploited   spherical symmetry to choose $\bX_j=\bx_{j0}:=(\sqrt{nP_j},0,\ldots, 0)$ and moved  all the constants to the right-hand-side.   
Let $\calE$ be the event $\{ \|\bx_{j0}+\bZ_j\|_2^2 = ns_j,\,\forall\, j  \in\{1,\ldots, k\} \}$. By introducing the independent random variables $\{U_j: j\in\calP^+\}$ that are related to $\{Z_{j1} : j\in\calP^+\}$ analogously to~\eqref{eqn:ZTheta}, we see that \eqref{eqn:Zk} reduces to 
\begin{equation}
h(s_1,\ldots, s_k;a,\mu)=\Pr\bigg(\sum_{j \in\calP^+}  \sqrt{P_js_j}\, U_{j }\in \Big[\frac{a_3}{n}, \frac{a_3+\mu}{n}\Big] \,\bigg|\, \calE\bigg), \label{eqn:reduce_inner_prod}
\end{equation}
where $a_3$ is related to $a_2$ by a  constant shift. In principle, since the $U_j$'s are independent, we can use its distribution in \eqref{eqn:densityU} to find the distribution of $\sum_{j \in\calP^+}  \sqrt{P_js_j}\, U_{j }$ by convolution and bound the probability using the steps that led to \eqref{eqn:final_bd}. However, the following method proves to be easier. Let $l$ be any element in $\calP^+$ then consider  
\begin{align}
&h(s_1,\ldots, s_k;a,\mu) \nn\\ 
& =\int  \Pr\bigg(\sum_{j \in\calP^+}   \sqrt{P_js_j} \,U_{j }\in \Big[\frac{a_3}{n}, \frac{a_3+\mu}{n}\Big] \,\bigg|\, \calE,\,\big\{\forall j \in\calP^+\setminus \{l\} , U_j=u_j\big\}\bigg)\,  \prod_{ j \in\calP^+\setminus \{l\} }f_{U_j|\calE}(u_j)\,  \rmd u_j   \label{eqn:law_tp}\\
& =\int   \Pr\bigg(   \sqrt{P_{l } s_{l }} \,U_{l  }\in \Big[\frac{a_4}{n}, \frac{a_4+\mu}{n}\Big] \,\bigg|\, \calE, \, \big\{\forall j \in\calP^+\setminus \{l\} , U_j=u_j\big\} \bigg)\,  \prod_{ j \in\calP^+\setminus\{l\} } f_{U_j|\calE}(u_j)\, \rmd u_j  \label{eqn:U_indep0}\\
 & =\int  \Pr\bigg(   \sqrt{P_{l } s_{l }}\, U_{l  }\in \Big[\frac{a_4}{n}, \frac{a_4+\mu}{n}\Big] \,\bigg|\, \calE \bigg)\,  \prod_{ j \in\calP^+\setminus \{l\} } f_{U_j|\calE}(u_j)\, \rmd u_j  \label{eqn:U_indep}\\
 & \le \int \frac{2\, L(P_{l },   s_{l }) \, \mu}{\sqrt{nP_{l }s_{l }}} \, \prod_{ j \in\calP^+\setminus \{l\}} f_{U_j|\calE}(u_j)\, \rmd u_j  \label{eqn:U_indep1}\\
 &= \frac{2\, L(P_{l },  s_{l }) \, \mu}{\sqrt{nP_{l }s_{l }}}   ,
\end{align}
where \eqref{eqn:law_tp} follows from the law of total probability; \eqref{eqn:U_indep0} follows by noting that $\{u_j:j\in\calP^+\setminus \{l\} \}$ are constants and defining $a_4$ to be related to $a_3$ by a constant shift;  \eqref{eqn:U_indep} is due to the joint independence of the random variables $\{U_j: j \in\calP^+\}$; and finally~\eqref{eqn:U_indep1}, which holds for $n$ sufficiently large, follows by the same reasoning in the steps that led to~\eqref{eqn:final_bd}.  Since $l\in\calP^+$ is arbitrary,
\begin{equation}
h(s_1,\ldots, s_k;a,\mu)\le\min_{l\in\calP^+} \frac{2\, L(P_{l } , s_{l }) \, \mu}{\sqrt{nP_{l }s_{l }}} .
\end{equation}
We conclude that,  just as in \eqref{eqn:root_n}, the probability $h(\by_1,\ldots, \by_k; a,\mu)$ is still  bounded above by a constant multiple of $\mu/\sqrt{n}$ and the constant does not depend on  $a$. The rest of the proof proceeds {\em mutatis mutandis}. 

\section{Proof of Lemma~\ref{lem:boundU}} 
\label{app:boundU}

We first find a lower bound for the normalization constant $F_n$ defined in \eqref{eqn:Fn}. Using the fact that $(1-u^2)^{-3/2} \geq 1$, we have 
\begin{equation}
F_n \ge \underline{F}_n:=\int_{-1}^1 \exp( n \alpha(u) ) \, \rmd u \label{eqn:underF}
\end{equation}
where the exponent is 
\begin{equation}
\alpha(u) :=  \frac{1}{2}\log (1-u^2)  + \sqrt{Ps} u .
\end{equation}
This exponent is maximized at 
\begin{equation}
u^* =  \frac{\sqrt{1+ 4Ps}-1}{2\sqrt{Ps}}, \label{eqn:u_star}
\end{equation}
which is in the interior of $[-1,1]$ for finite $P$. Furthermore, the second derivative of $\alpha$  is 
\begin{equation}
\alpha''(u) = -\frac{(1+u^2)}{(1-u^2)^2}
\end{equation}
which is always negative. Now we use  Laplace's method to lower bound the definite integral in \eqref{eqn:underF} with that of a  Gaussian~\cite{Tierney86, Shun95}. We provide the details for the reader's convenience. Let $\epsilon\in(0, -\alpha''(u^*))$.  By the continuity of $\alpha''(u)$ at $u^*$ and Taylor's theorem, there exists a $\zeta \in (0, 1-u^*)$ such that for any $u \in (u^*-\zeta,u^*+\zeta)\subset [-1, 1]$, we have $\alpha(u)\ge \alpha(u^*)+\frac{1}{2}(\alpha''(u^*)-\epsilon)(u-u^*)^2$.   The following lower bounds hold:
\begin{align}
 \underline{F}_n  &\ge\int_{u^*-\zeta}^{u^*+\zeta} \exp(n\alpha(u))\, \rmd u \\
 &\ge \exp(n\alpha(u^*))\int_{u^*-\zeta}^{u^*+\zeta} \exp\Big( \frac{n}{2}(\alpha''(u^*)-\epsilon) (u-u^*)^2\Big)\, \rmd u \\
 &=\exp(n\alpha(u^*)) \sqrt{ \frac{1}{n( -\alpha''(u^*)+\epsilon)}} \int_{ -\zeta\sqrt{ n ( -\alpha''(u^*)+\epsilon) }}^{ \zeta\sqrt{ n ( -\alpha''(u^*)-\epsilon)  }}   \rme^{-v^2/2} \, \rmd v . \label{eqn:change_vars}
\end{align}
We used the change of variables $v=\sqrt{n ( -\alpha''(u^*)+\epsilon)}(u-u^*)$ in the final step. The integral in \eqref{eqn:change_vars} tends to $\sqrt{2\pi}$ as $n$ becomes large so 
\begin{equation}
\liminf_{n\to\infty}\frac{\underline{F}_n}{ \sqrt{ \frac{2\pi}{n  |\alpha''(u^*)|}} \exp(n\alpha(u^*))}\ge \sqrt{\frac{- \alpha''(u^*) }{-\alpha''(u^*)+\epsilon }} . \label{eqn:take_liminf}
\end{equation}
Since $\epsilon>0$ is arbitrary, we can rewrite \eqref{eqn:take_liminf} as
\begin{equation}
\underline{F}_n \ge \gamma_n \, \sqrt{\frac{2\pi}{n |\alpha''(u^*)|}}\exp(n \alpha(u^*)), \label{eqn:laplace}
\end{equation}
for some sequence $\gamma_n$  that converges to $1$ as $n \to \infty$.    Furthermore, the numerator of  $f_{U|\calE}(u)$ in \eqref{eqn:densityU} can be upper bounded as 
\begin{equation}
(1-u^2)^{(n-3)/2} \exp\big( n \sqrt{Ps} u \big) = \exp (n \beta_n(u)) \le\exp( n \beta_n(u_n^*)) \label{eqn:upper_bd_num}
\end{equation}
where the exponent is 
\begin{equation}
\beta_n(u):= \Big(\frac{1}{2}-\frac{3}{2n}\Big)\log (1-u^2)  + \sqrt{Ps} u 
\end{equation}
and the maximizer of $\beta_n(u)$ is 
\begin{equation}
u_n^* :=    \frac{   \sqrt{  (1-\frac{3}{n} )^2 + 4Ps  }-(1-\frac{3}{n})}{2 \sqrt{Ps}}.
\end{equation} 
Clearly, $u_n^*\to u^*$ as $n\to\infty$. We have,  by uniting \eqref{eqn:laplace} and \eqref{eqn:upper_bd_num}, that
\begin{equation}
\sup_{u \in [-1, 1]} f_{U|\calE}(u) \le \frac1{\gamma_n} \sqrt{\frac{n |\alpha''(u^*)|}{2\pi}} \exp\big( n [\beta_n(u_n^*)-\alpha(u^*)   ] \big) .\label{eqn:sup_u}
\end{equation}
Now, we examine the exponent $\beta_n(u_n^*)-\alpha(u^*)$ above. We have 
\begin{align}
  \beta_n(u_n^*)-\alpha(u^*)   \le \beta_n(u_n^*)-\alpha(u_n^* )  =  \frac{3}{2n}\log\frac{1}{1-(u_n^*)^2}   \label{eqn:def_alpha_beta} 
\end{align}
where the inequality  follows because $u^*$ maximizes $\alpha$ and so $\alpha(u_n^*)\le\alpha(u^*)$ and  the equality is due to the definitions of $\alpha(u)$ and $\beta_n(u)$. Thus,  \eqref{eqn:sup_u} can be further upper bounded as 
\begin{equation}
\sup_{u \in [-1, 1]} f_{U|\calE}(u) \le \frac1{\gamma_n} \cdot\sqrt{\frac{n |\alpha''(u^*)|}{2\pi}} \cdot\frac{1}{(1-(u_n^*)^2)^{3/2}}   .
\end{equation}
Dividing both sides  by $\sqrt{n}$ and taking the $\limsup$ shows that the upper bound can be chosen to be 
\begin{equation}
L(P,s) =  \frac{1}{ (1-(u^*)^2)^{3/2}} \cdot\sqrt{\frac{  |\alpha''(u^*)|}{2\pi}}  =  \sqrt{ \frac{1+ (u^*)^2}{ 2\pi (1-(u^*)^2)^5} }\label{eqn:defB}  .
\end{equation}
This concurs with \eqref{eqn:defL} after we substitute for the  value of $u^*$ in \eqref{eqn:u_star}.\qed

\subsection*{Acknowledgements}
VT sincerely thanks Shaowei Lin (I$^2$R, A*STAR) for many helpful explanations concerning approximation of integrals in high dimensions.  The authors also thank  Jonathan Scarlett (Cambridge) and  Y\"ucel Altu\u{g} (Cornell) for  discussions and constructive comments on the manuscript.
\bibliographystyle{ieeetr}
\bibliography{../isitbib}

\begin{thebibliography}{10}

\bibitem{Shannon48}
C.~E. Shannon, ``A mathematical theory of communication,'' {\em Bell Systems
  Technical Journal}, vol.~27, pp.~379--423, 1948.

\bibitem{Sha59b}
C.~E. Shannon, ``Probability of error for optimal codes in a {G}aussian
  channel,'' {\em Bell Systems Technical Journal}, vol.~38, pp.~611--656, 1959.

\bibitem{Yoshihara}
K.~Yoshihara, ``Simple proofs for the strong converse theorems in some
  channels,'' {\em Kodai Mathematical Journal}, vol.~16, no.~4, pp.~213--222,
  1964.

\bibitem{Wolfowitz}
J.~Wolfowitz, {\em Coding Theorems of Information Theory}.
\newblock Springer-Verlag, New York, 3rd~ed., 1978.

\bibitem{Strassen}
V.~Strassen, ``{Asymptotische Absch\"{a}tzungen in Shannons
  Informationstheorie},'' in {\em Trans. Third Prague Conf. Inf. Theory},
  (Prague), pp.~689--723, 1962.

\bibitem{Hayashi09}
M.~Hayashi, ``Information spectrum approach to second-order coding rate in
  channel coding,'' {\em IEEE Trans. on Inf. Th.}, vol.~55, pp.~4947--4966, Nov
  2009.

\bibitem{PPV10}
Y.~Polyanskiy, H.~V. Poor, and S.~Verd\'{u}, ``Channel coding rate in the
  finite blocklength regime,'' {\em IEEE Trans. on Inf. Th.}, vol.~56,
  pp.~2307--2359, May 2010.

\bibitem{TomTan12}
M.~Tomamichel and V.~Y.~F. Tan, ``A tight upper bound for the third-order
  asymptotics of most discrete memoryless channels,'' {\em IEEE Trans. on Inf.
  Th.}, vol.~59, pp.~7041--7051, Nov 2013.

\bibitem{Pol10}
Y.~Polyanskiy, {\em Channel coding: {Non}-asymptotic fundamental limits}.
\newblock PhD thesis, Princeton University, 2010.

\bibitem{altug13}
Y.~Altu\u{g} and A.~B. Wagner, ``The third-order term in the normal
  approximation for singular channels,'' {\em {\tt arXiv:1309.5126 [cs.IT]}},
  Sep 2013.

\bibitem{mou13b}
P.~Moulin, ``The log-volume of optimal codes for memoryless channels, within a
  few nats,'' {\em {\tt arXiv:1311.0181 [cs.IT]}}, Nov 2013.

\bibitem{Cov06}
T.~M. Cover and J.~A. Thomas, {\em Elements of Information Theory}.
\newblock Wiley-Interscience, 2nd~ed., 2006.

\bibitem{gallagerIT}
R.~G. Gallager, {\em {Information Theory and Reliable Communication}}.
\newblock New York: Wiley, 1968.

\bibitem{altug11}
Y.~Altu\u{g} and A.~B. Wagner, ``Refinement of the sphere packing bound for
  symmetric channels,'' in {\em Proc. 49th Annual Allerton Conf. Communication,
  Control, and Computing}, 2011.

\bibitem{altug12}
Y.~Altu\u{g} and A.~B. Wagner, ``A refinement of the random coding bound,'' in
  {\em Proc. 50th Annual Allerton Conf. Communication, Control, and Computing},
  2012.

\bibitem{altug12a}
Y.~Altu\u{g} and A.~B. Wagner, ``Refinement of the sphere packing bound,'' in
  {\em Int. Symp. Inf. Th.}, (Cambridge, MA), 2012.

\bibitem{Sca13}
J.~Scarlett, A.~Martinez, and A.~{Guill\'{e}n i F\`{a}bregas}, ``A derivation
  of the asymptotic random-coding prefactor,'' in {\em Proc. 51st Annual
  Allerton Conf. Communication, Control, and Computing}, 2013.
\newblock {\tt arXiv:1306.6203 [cs.IT]}.

\bibitem{Pol13}
Y.~Polyanskiy, ``Saddle point in the minimax converse for channel coding,''
  {\em IEEE Trans. on Inf. Th.}, vol.~59, pp.~2576--2595, May 2013.

\bibitem{Dembo}
A.~Dembo and O.~Zeitouni, {\em Large Deviations Techniques and Applications}.
\newblock Springer, 2nd~ed., 1998.

\bibitem{Stam}
A.~J. Stam, ``Limit theorems for uniform distributions on spheres in
  high-dimensional {E}uclidean spaces,'' {\em Journal of Applied Probability},
  vol.~19, no.~1, pp.~221--228, 1982.

\bibitem{Mol13}
E.~{MolavianJazi} and J.~N. Laneman, ``A finite-blocklength perspective on
  {G}aussian multi-access channels,'' {\em {\tt arXiv:1309.2343 [cs.IT]}}, Sep
  2013.

\bibitem{ScarlettTan}
J.~Scarlett and V.~Y.~F. Tan, ``Second-order asymptotics for the {Gaussian MAC}
  with degraded message sets,'' {\em {\tt arXiv:1310.1197 [cs.IT]}}, Oct 2013.

\bibitem{feller}
W.~Feller, {\em An Introduction to Probability Theory and Its Applications}.
\newblock John Wiley and Sons, 2nd~ed., 1971.

\bibitem{Tierney86}
L.~Tierney and J.~B. Kadane, ``Accurate approximations for posterior moments
  and marginal densities,'' {\em Journal of the American Statistical
  Association}, vol.~81, pp.~82--86, Mar 1986.

\bibitem{Shun95}
Z.~Shun and P.~McCullagh, ``Laplace approximation of high dimensional
  integrals,'' {\em Journal of the Royal Statistical Society, Series B
  (Methodology)}, vol.~57, no.~4, pp.~749--760, 1995.

\end{thebibliography}

\end{document}